\documentclass{iaus}
\usepackage{graphicx}

\newcommand{\Msun}{\mbox{$M_{\odot}$}}

\title[Observational studies of mass loss from AGB stars] %% give here short title %%
{Observational studies of mass loss from AGB stars}

\author[M.~Matsuura]   %% give here short author list %%
{Mikako~Matsuura$^{1}$
}
\affiliation{$^{1}$Department of Physics and Astronomy,
University College London, 
Gower Street, London WC1E 6BT,
United Kingdom\\email: {\tt  mikako@star.ucl.ac.uk} \\}

\pubyear{2012}
\volume{283}  %% insert here IAU Symposium No.
\pagerange{1--4}
\jname{Planetary Nebulae: an Eye to the Future}
\editors{A.C. Editor, B.D. Editor \& C.E. Editor, eds.}
\begin{document}

\maketitle

\begin{abstract}

It is important to properly describe the mass-loss rate of AGB stars, in order to understand their evolution from the AGB to PN phase. 
The primary goal of this study is to investigate the influence of metallicity on the mass-loss rate,
under well determined luminosities.

The luminosity of the star is a crucial parameter for the radiative driven stellar wind. Many efforts have been invested to constrain the AGB mass-loss rate, but most of the previous studies use Galactic objects, which have poorly known distances, thus their luminosities. To overcome this problem, we have studied mass loss from AGB stars in the Galaxies of the Local Group. The distance to the stars have been independently measured, thus AGB stars in these galaxies are ideal for understanding the mass-loss rate. Moreover, these galaxies have a lower metallicity than the Milky Way, providing an ideal target to study the influence of metallicity on the mass-loss rate.

We report our analysis of mass loss, using the Spitzer Space Telescope and the Herschel Space Observatory.
We will discuss the influence of AGB mass-loss on stellar evolution, and explore AGB and PN contribution to the lifecycle of matter in galaxies. 
\end{abstract}

\firstsection % if your document starts with a section,
\section{Introduction}

Spitzer Space Telescope (\cite[Werner et al. 2004]{Werner04}) has opened up a new frontier of studying asymptotic giant branch (AGB) stars beyond the Milky Way. This includes the photometric surveys of the Magellanic Clouds (MCs; \cite[Meixner et al. 2006; Gordon et al. 2011]{Meixner06, Gordon11}) and spectroscopic studies of selected AGB stars in these galaxies (\cite[Zijlstra et al. 2006; Sloan et al. 2006; Buchanan et al. 2006; Lagadec et al. 2007; Kemper et al. 2010]{Zijlstra06, Sloan06, Buchanan06, Lagadec07, Kemper10}), and the dwarf spheroidal galaxies (\cite[Matsuura et al. 2007; Sloan et al. 2009]{Matsuura07, Sloan09}).
Because distances to AGB stars can be determined independently to AGB stars, we can easily estimate luminosities of AGB stars, which is the largest uncertainty parameter to drive the mass-loss rate for Galactic objects.
Further, these galaxies have lower metallicities than the Milky Way, and they are ideal to study how metallicity influences dust composition and mass-loss rate of AGB stars.
The typical metallicities of the Large and Small Magellanic Clouds are
half and quarter of the Solar metallicity, respectively (\cite[Russell \& Bessell 1989)]{Russell89}.
Here we briefly summarise our resent progress of our studies of AGB stars.

\section{Analysis and results of extra-galactic AGB stars}

We have taken the photometric data of the Magellanic Clouds from  \cite{Meixner06} and \cite{Gordon11} and these were investigated by these many consortia \cite[(Blum  et al. 2006, Boyer et al. 2011)]{Blum06, Boyer11}.
We have cross-identified the LMC photometric data with spectroscopically known objects in the LMC, including AGB stars, HII regions, YSOs and post-AGB stars (\cite[Matsuura et al. 2009]{Matsuura09}).
Recent up-dates by \cite{Woods11} and \cite{Ita08} will be incorporated in future work.
The resultant colour-colour diagram of the objects in the Large Magellanic Cloud (LMC)
is plotted in Figure 1.
Sequences of mass-loss rate estimated from \cite{Groenewegen07}, and 
correlation derived from  \cite{Groenewegen07} are overlaid
in dotted lines. The redder colour corresponds to higher mass-loss rate.

Figure 2 shows the same combination of the
colour-colour diagram, but for the Small Magellanic Cloud (SMC).
There is a slight difference in the distributions of stars, compared with LMC diagram:
the SMC has very few oxygen-rich AGB stars with colour redder than $K-[8.0]>1.8$, 
which corresponds to approximately a mass-loss rate higher than $>10^{-6}$\,\Msun\,yr$^{-1}$.
Near-infrared imaging surveys of the LMC and the SMC, showed that the number ratio of oxygen-rich stars to carbon-rich stars decreases (\cite[Richer \& Westerlund 1983]{Richer83}), which have similar metallicities, going from the Galaxy (about the Solar), 
to the LMC (about 1/2 of the Solar metallicity), to the SMC (about 1/4 of the solar metallicity).
We found that the impact of the metallicities is not only on the number ratio of oxygen-rich and carbon-rich, but also on the dust formation, and mass-loss rates.
After the 3rd-dredge ups,
AGB stars might turn from oxygen-rich to carbon-rich,
and an circumstellar envelope could have a sufficient number of carbon atoms for the formation of carbonaceous dust grains.
Radiative pressure on the dust grains drives the mass loss from AGB stars, thus, mass-loss rates of carbon-rich AGB stars do not reduce significantly towards lower metallicities, at least down to quarter of the solar metallicity.
This is consistent with earlier studies based on smaller number of spectroscopic sample
(e.g. \cite[Matsuura et al. 2005]{Matsuura05}) and theoretical studies (\cite[Wachter et al. 2008]{Wachter08}).
%It is still unknown other effects, such as energy injection from pulsation
%could be important, even further lower metallicity \cite{Mattson}
However, in oxygen-rich AGB stars, the dominant species of dust grains are silicate, and their mass is mainly limited by the silicon elemental abundance.
Silicon atoms are not synthesised inside AGB stars, and their abundance is limited by silicon incorporated into the star when the stars were formed.
Silicate abundances in AGB stars largely correlate with the metallicities of the parent galaxies.
Thus, the mass-loss rate of oxygen-rich AGB stars reduces towards lower metallicity, as theoretical prediction by \cite{Bowen91} showed.
Hence, it is more difficult for oxygen-rich stars to form significant amounts of dust to drive a high mass-loss rate.

Figure 2 shows that the number ratio decrease even drastically towards higher mass-loss rate stars in the SMC. 
As the LMC and SMC have experienced more or less similar star-formation histories
(\cite[Zaritsky \& Harris 2004; Harris \& Zaritsky 2009]{Zaritsky04, Harris09}), 
the age of stars would not be the main factor to explain this change in number ratio.
This is probably caused by the effect of metallicities. 
It appears that AGB stars change from from oxygen-rich to carbon-rich, before they reach a high mass-loss rate, suggesting that the turning from oxygen-rich to carbon-rich might have happened at an earlier age during the AGB stellar evolution.
Note that the cluster of point sources near K$-$[8.0]$\sim$2 and K$-$[24]$\sim$6 are external galaxies.

From these mass-loss rates of individual AGB stars, we can estimate the total gas and dust inputs from AGB stars into the interstellar medium (ISM) of the LMC and the SMC.
The estimated total gas inputs is 2--4$\times10^{-2}$\,\Msun\,yr$^{-1}$ for LMC adding both oxygen-rich and carbon-rich, and $\sim$0.2$\times10^{-2}$\,\Msun\,yr$^{-1}$ for the SMC.
The estimate particularly for the SMC suffers from contamination of external galaxies, and this requires a future revision and elaborate removal data from these galaxies.
The gas inputs from type II supernovae are 2--4$\times10^{-2}$\,\Msun\,yr$^{-1}$ for the LMC, and 0.8$\times10^{-2}$\,\Msun\,yr$^{-1}$ for the SMC.
This estimate is based on the SN rates in these galaxies \cite[(Maoz \& Badenes 2010)]{Maoz10}.
In the LMC, both AGB stars and SNe contribute comparably to the gas inputs, while in the SMC, AGB stars appear to be smaller inputs of gas into the ISM than SNe. \cite{Matsuura09} found that dust inputs from AGB stars are small, compared with dust present in the ISM of the LMC.
Recent finding of dust in SN 1987A (\cite[Matsuura et al. 2011]{Matsuura11}) and from the Herschel Magellanic Survey 
(\cite[Meixner et al. 2010]{Meixner10}) suggests dust inputs from supernovae into the ISM as important contributors.

The mass-loss study of AGB stars continue with Herschel Observations, using the data from the Herschel guaranteed time observations (\cite[Groenewegen et al. 2011]{Groenewegen11}), which now started revealing a complex structure 
of the atmosphere (\cite[Decin et al. 2010; Royer et al. 2010; Wesson et al. 2010]{Decin10, Royer10, Wesson10}).
We are working on modelling spectra using the non-LTE line radiative transfer code, SMMOL \cite[(Rawlings et al. 2001)]{Rawlings01}.

%%%%%%%
\begin{figure}[b]
% \vspace*{-2.0 cm}
\begin{center}
    \resizebox{\hsize}{!}{\includegraphics[angle=90]{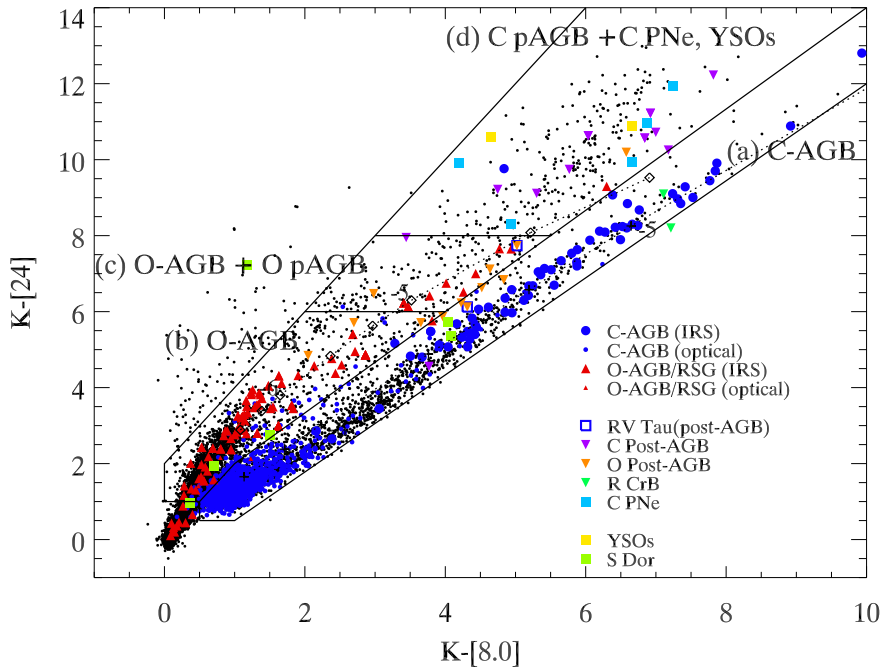}} 
 \vspace*{-0.8 cm}
 \caption{Colour-colour diagramme of point sources in the Large Magellanic Cloud.
 The objects with spectroscopic identified classifications are overplotted.
 The sequence of oxygen-rich objects and carbon-rich objects are well separated.
 Two dotted lines show the estimated mass-loss rate and colour relations for oxygen-rich
 and carbon-rich stars, respectively.}
   \label{fig1}
\end{center}
\end{figure}
%%%%%%%

%%%%%%%
\begin{figure}[b]
% \vspace*{-2.0 cm}
\begin{center}
    \resizebox{\hsize}{!}{\includegraphics[angle=90]{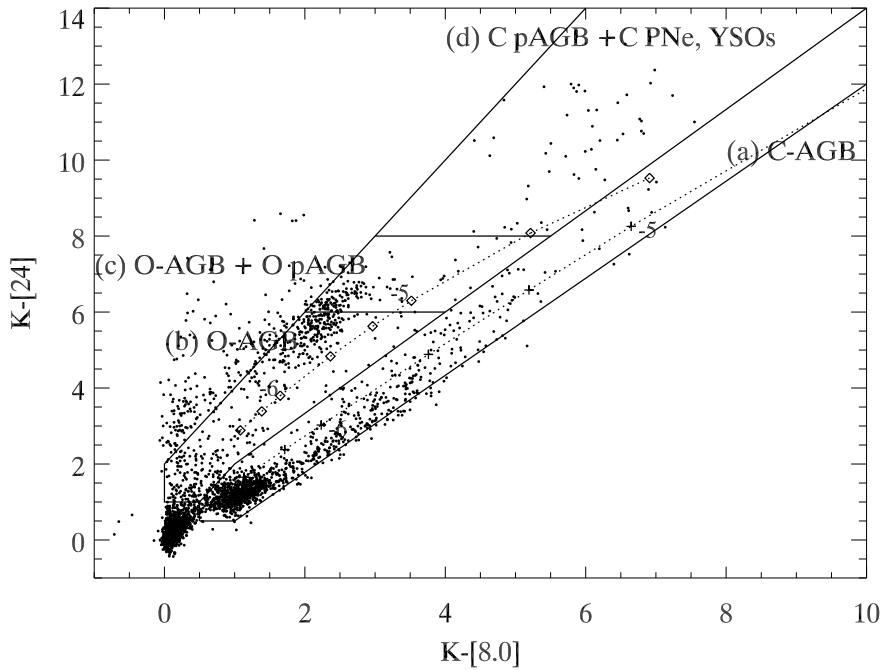}} 
 \vspace*{-0.8 cm}
 \caption{Same as figure 1, but for the objects towards Small Magellanic Cloud (SMC).
 There are very few number of oxygen-rich stars with colour redder than $K-[8.0]>1.8$, 
 which corresponds to approximately mass-loss rate
 higher than $>10^{-6}$\,\Msun\,yr\,s$^{-1}$. 
 Note that the cluster of point sources near K$-$[8.0]$\sim$2 and K$-$[24]$\sim$6 are
 external galaxies located behind the SMC.}
   \label{fig2}
\end{center}
\end{figure}
%%%%%%%

\bibliography{references}{}

\end{document}